\documentstyle[12pt]{article}

\catcode`\@=11
\def\numberbysection{\@addtoreset{equation}{section}
	\def\theequation{\thesection.\arabic{equation}}}
\def\beq{\begin{equation}}
\def\eeq{\end{equation}}
\def\barr{\begin{eqnarray}}
\def\earr{\end{eqnarray}}
\def\winf{W_{1+\infty}\ }
\def\u1{\widehat{U(1)}}
\def\rep{ representation }
\def\reps{ representations }

\numberbysection
\begin{document}
\begin{titlepage}
\begin{center}
\hfill DFTT 10/95 \\
\vskip .3 in
{\large \bf Quantum Hall Fluids as \\
$\winf$ Minimal Models}\footnote{To appear in the proceedings of the
conference on {\it Statistical Mechanics and Quantum Field Theory},
University of Southern California, Los Angeles (U.S.A.), May 1994 and
of the summer school on {\it Particles, Fields and Strings},
Banff (Canada), August 1994.}
\vskip 0.2in
Andrea CAPPELLI \\
{\em I.N.F.N. and Dipartimento di Fisica, Largo E. Fermi 2,
 I-50125 Firenze, Italy}
\\
\vskip 0.1in
Carlo~A.~TRUGENBERGER\footnote{Supported by a Profil 2 fellowship of
the Swiss National Science Foundation.}  \\
{\em D\'epartement de Physique Th\'eorique, Univ. de Gen\`eve,
24 quai E. Ansermet, CH-1211, Gen\`eve 4, Switzerland}
\\
\vskip 0.1in
Guillermo~R.~ZEMBA \\
{\em I.N.F.N. and Dipartimento di Fisica Teorica,
	  Via P. Giuria 1, I-10125 Torino, Italy}
\end{center}
\vskip .1 in
\begin{abstract}
We review our recent work on the algebraic characterization of
quantum Hall fluids. Specifically, we explain how the incompressible
quantum fluid ground states can be classified by effective edge
field theories with the $W_{1+\infty }$ dynamical symmetry of
``quantum area-preserving diffeomorphisms''. Using the representation
theory of $W_{1+\infty }$, we show how all fluids with filling
factors $\nu = m/(pm +1)$ and $\nu = m/(pm-1)$ with $m$ and $p$
positive integers, $p$ even, correspond exactly to the $W_{1+\infty }$
{\it minimal models}.
\end{abstract}
\vfill
February 1995\hfill\\
\end{titlepage}
\pagenumbering{arabic}
%

\section{Introduction}

The quantum Hall effect \cite{prange}
provides fascinating examples of {\it quantum
fluids}. At low temperatures, interacting planar electrons in high
magnetic fields $B$ have strong quantum correlations which lead to
collective motion and macroscopic quantum effects. These find their
experimental evidence in a discrete series of plateaus at rational
values of the Hall conductivity:
\barr
\sigma _{xy}& = &\ { e^2\over h}\ \nu\ ,\nonumber\\
\nu & = &\ 1, {1\over 3}, {1\over 5},{2\over 7},\dots,2,\dots .
\label{cond}\earr
Corresponding to these plateaus the longitudinal
conductivity $\sigma _{xx}$ vanishes.
The same plateaus are observed in several materials, signalling
{\it universality}.
Another experimental result is
the remarkable {\it exactness} of these rational values of $\nu $;
the experimental error is $\Delta\nu=10^{-8}$ for integer $\nu$.

The current understanding of the quantum Hall effect is based
on the seminal work of Laughlin \cite{laugh}.
The main idea is the
existence of {\it incompressible quantum fluids} at specific rational
values $\bar\rho =\nu B/2\pi $ ($\hbar=1$, $c=1$)
of the electron density. These are very stable,
macroscopic quantum states with uniform density and an energy gap.
Incompressibility
accounts for the lack of low-lying conduction modes, which causes
$\sigma _{xx}$ to vanish, while the Hall conduction is realized as an
overall rigid motion of the uniform droplet, which gives eq. (\ref{cond}).

While Laughlin's theory is very successful, the observed exactness and
universality calls for a {\it fundamental principle} underlying it.
Indeed, the {\it universality} observed in experiments calls for an effective
theory approach at long distances,
while the extreme precision of the rational values of $\nu$ suggests
that dynamics is constrained by {\it symmetry}.
Both facts suggest an analogy with two-dimensional critical phenomena, which
are classified by conformal field theories \cite{bpz}.

The effective field theory approach, developed by Landau, Ginsburg,
Wilson and others \cite{polch}, does not attempt to solve
the microscopic many-body dynamics, but rather it guesses the
macroscopic physics generated by this dynamics.
The variables of the effective field theory are the relevant low-energy
(long-distance) degrees of freedom, which are characterized by a specific
symmetry.
They describe universal properties, which are independent of the
microscopic details.
This approach is well suited for the quantum Hall effect, given the
 very precise and universal values of the Hall conductivity.

In the following, we shall present an overview of our
\cite{ctz1}\cite{cdtz1}\cite{cdtz2}\cite{ctz2}\cite{ctz3}\cite{ctz4}\cite{ctz5}
and related \cite{sakita}\cite{flohr} recent work
on the effective field theory approach to the quantum Hall effect.
This is based on an {\it algebraic characterization} of incompressible
quantum fluids.


\section{Dynamical symmetry and kinematics of incompressible fluids}

In this section, we shall review the dynamical symmetry characterizing
{\it chiral}, two-dimensional, incompressible quantum fluids and indicate
how this leads uniquely to the construction of the Hilbert spaces
of low-energy excitations.

\bigskip
\noindent{\bf Classical fluids}

A classical incompressible fluid is defined by its distribution function
\beq
\rho (z, \bar z, t) = \rho_0 \ \chi _{S_A(t)}\ ,\qquad
\rho_0 \equiv {N\over A}\ ,
\label{rho}\eeq
where $\chi_{S_A(t)}$ is the characteristic function for a surface
$S_A(t)$ of area $A$, and $z=x+iy$, $\bar z= x-iy$ are complex
coordinates on the plane.
Since the particle number $N$ and the average
density $\rho_0$ are constant, the area $A$ is also {\it constant}.
The only possible change in response to external forces is in the
shape of the surface.
The shape changes at constant area can be generated by
{\it area-preserving diffeomorphisms} of the two-dimensional plane.
Thus, the configuration space of a classical incompressible fluid
can be generated by applying these transformations to a reference
droplet.

Next we recall the Liouville theorem,
which states that canonical transformations preserve the
phase-space volume. Area-preserving diffeomorphisms are, therefore,
canonical transformations of a two-dimensional phase space.
In order to use the formalism of canonical transformations, we
treat the original coordinate plane as a {\it phase space}, by
postulating non-vanishing Poisson brackets between $z$ and $\bar z$.
We do this by defining the dimensionless Poisson brackets
\beq
\{f, g\} \equiv {i\over \rho_0} \ \left( \partial f
\bar \partial g - \bar \partial f \partial g \right) \ , \label{pob} \eeq
where $\partial \equiv \partial / \partial z\ $ and
$\ \bar \partial \equiv \partial /\partial \bar z$, so that
\beq
\{ z, \bar z\}= {i\over \rho_0}\ .\label{cmo} \eeq
Note that the Poisson brackets select a preferred {\it chirality},
because they are
not invariant under the two-dimensional parity transformation
$z\to \bar z,\ \bar z\to z$; in the
quantum Hall effect, the parity breaking
is due to the external magnetic field.

Area-preserving diffeomorphisms, {\it i.e.}, canonical transformations,
are usually defined in terms of a generating function
${\cal L}(z,\bar z)$ of both ``coordinate'' and ``momentum'',
as follows:
\beq
\delta z\ =\ \{{\cal L}, z\}\ ,\qquad
\delta {\bar z}\ =\ \{{\cal L}, {\bar z}\}\ . \label{wdef} \eeq
A basis of (dimensionless) generators is given by
\beq
{\cal L}^{(cl)}_{n,m}\equiv \rho_0 ^{{n+m}\over 2} \ z^n
\bar z^m \ .\label{gen} \eeq
These satisfy the classical $w_{\infty }$ algebra \cite{shen}
\beq
\left\{ {\cal L}^{(cl)}_{n,m}, {\cal L}^{(cl)}_{k,l} \right\}
=-i\ (mk-nl)\ {\cal L}^{(cl)}_{n+k-1,m+l-1} \ . \label{clw} \eeq

Let us now discuss how $w_{\infty }$ transformations can be used
to generate the configuration space of classical excitations
above the ground state.
These configurations have a classical energy due to the inter-particle
interaction and the external confining potential, whose specific form
is not needed here.
Let us assume a generic convex and rotation-invariant energy function,
such that the minimal energy configuration $\rho_{GS}$
has the shape of a disk of radius $R$:
\beq
\rho_{GS}(z, \bar z)=\rho _0 \ \Theta \left( R^2-z\bar z
\right) \ ,\label{gsd} \eeq
where $\Theta$ is the Heaviside step function.
The classical "small excitations" around this ground state
configuration are given by the infinitesimal deformations
of $\rho_{GS}$ under area-preserving diffeomorphisms,
\beq
\delta \rho _{n,m} \equiv \left\{ {\cal L}^{(cl)}_{n,m},
\rho _{GS} \right\} \ .\label{sex} \eeq
Using the Poisson brackets (\ref{pob}) , we obtain
\beq
\delta \rho _{n,m} = i \left( \rho _0 R^2 \right) ^{{n+m}\over 2}
(m-n)\ {\rm e}^{i(n-m)\theta }\ \delta \left( R^2 -z\bar z\right) \ .
\label{wav} \eeq
These correspond to density fluctuations localized on the
sharp boundary
(which is parametrized by the angle $\theta$) of the classical
droplet. Due to the dynamics provided by the energy
function, they will propagate on the boundary
with a frequency $\omega _k$ dependent on the angular momentum
$k\equiv (n-m)$, thereby turning into {\it edge waves}.
These are the eigenoscillations of the classical incompressible
fluid.

Another type of excitations are classical vortices in the bulk of
the droplet, which correspond to localized holes or dips in the density.
The absence of density waves, due to incompressibility, implies that
any localized density excess or defect is transmitted completely
to the boundary, where it is seen as a further edge deformation.
For each given vorticity in the bulk, we can then construct the
corresponding
basis of edge waves in a fashion analogous to (\ref{sex}) . Thus, the
configuration space of the excitations of a classical incompressible
fluid (of a given vorticity) is spanned by infinitesimal $w_{\infty }$
transformations. This is the {\it dynamical symmetry}
of classical incompressible fluids.

\bigskip
\noindent{\bf Quantum fluids and their edge excitations}

The quantum \footnote{
Throughout this paper we shall use units such that $c=1$, $\hbar =1$.}
version of the chiral, incompressible fluids is given by the
Laughlin theory of the plateaus of the quantum Hall effect \cite{laugh} .
The simplest example of such a macroscopic quantum state is a
fully filled Landau level (filling fraction $\nu =1$).
Generically, it possesses three types of excitations.
First, there are {\it gapless edge excitations} \cite{wen} , which
are the quantum descendants of the classical
edge waves described before. These
are particle-hole excitations across the Fermi surface represented
by the edge of the droplet of radius $R$ \cite{stone}.
They are {\it gapless} because their energy, of $O(1/R)$,
vanishes for $R\to\infty$.
Second, there are localized
quasi-particle and quasi-hole excitations, which have a finite gap. These
are the quantum analogs of the classical vortices and correspond to the
anyon excitations \cite{laugh} with fractional charge, spin and statistics
\cite{wilc}.
As in the classical case, they manifest themselves as charged excitations
at the edge, owing to incompressibility.
The third type of excitations are two-dimensional density waves in the
bulk, the magnetoplasmons and (for $\nu <1$) the magnetophonons
\cite{gmp}.
These have higher gaps and are not included in our
effective field theory approach.

In the previous section, we have explained the connection between the
classical edge waves and the generators of the algebra $w_{\infty }$
of area-preserving diffeomorphisms.
In the quantum theory, there is a corresponding relation
between edge excitations and
the generators of the quantum version of $w_{\infty }$, called
$W_{1+\infty }$ \cite{shen} .
This algebra is obtained by replacing the Poisson brackets
ref{pob} with quantum commutators: $i\{\ ,\ \}\ \to\ [\ ,\ ]\ $,
and by taking the thermodynamic limit \cite{cdtz1} .

In this limit, the radius of the droplet grows as
$R\propto \ell\sqrt{N}$, where $\ell=\sqrt{2/(eB)}$ is the
magnetic length and $B$ the magnetic field.
Quantum edge excitations, instead, are confined
to a boundary annulus of finite size $O(\ell )$.
In the $N \to \infty $ limit, therefore, edge excitations become the
particle hole excitations of a relativistic theory describing a
Weyl (chiral) fermion living on the one-dimensional edge of the droplet
\cite{cdtz1} . In this limit, the quantum incompressible fluid becomes the
Dirac sea for this relativistic theory. Charged fermions represent instead
quasi-particle excitations.

The field operator for the Weyl fermion
\footnote{
Hereafter, we choose units such that $\ell = 1$.} is given by \cite{bpz}
\beq
F_R(\theta)\ =\ {1\over \sqrt{ R}}\
\sum_{k=-\infty}^{\infty}\ {\rm e}^{i(k-1/2)\theta}\ b_k\ ,
\qquad \left( |z|=R\ ,\ t=0\right)\ ,\label{weyl} \eeq
where $\theta$ parametrizes the circular boundary,
$b_k$ and $b^{\dag }_k$ are fermionic Fock space operators satisfying
$\left\{ b_l, b^{\dag }_k \right\} = \delta _{l,k} \ $,
and $k$ is the angular momentum measured with respect to the ground
state value.

The generators of the quantum algebra $\winf$ are represented
in this Fock space by the bilinears
\barr
V_n^j\ & = &\ \int_0^{2\pi} {d\theta \over{2\pi}}\ :\
F^{\dagger}(\theta)\ {\rm e}^{-in\theta}\ g^j_n \left(i\partial_{\theta}
\right)\ F(\theta)\ : \nonumber\\
\ & = &\ \sum_{k=-\infty}^{\infty}\ p(k,n,j)\ : b^{\dag}_{k-n}\ b_k : \ ,
\qquad j\ge 0\ . \label{acf} \earr
In this expression,
$F(\theta) = F_R (\theta)\ {\rm e}^{i\theta /2}\sqrt{R}$
is the canonical form of the Weyl field operator of conformal field theory.
The factor $\ g^j_n \left(i\partial_{\theta}\right)\ $
is a $j$-th order polynomial in $i\partial_{\theta}$, whose form
specifies the basis of operators and guarantees the hermiticity
$\left( V^j_n \right)^{\dag} = V^j_{-n}$. The
coefficients $p(k,n,j)$ are also $j$-th order polynomials in $k$
which we do not need to specify here (see \cite{ctz4}).
The $\winf$ algebra reads
\beq
\left[\ V^i_n, V^j_m\ \right] =(jn-im) \ V^{i+j-1}_{n+m} +
q(i,j,m,n) \ V^{i+j-3}_{n+m} +\dots +  c^i(n) \ \delta ^{i,j}
\delta _{n+m,0} \ . \label{Win} \eeq
Here, $i+1=h \ge 1$ represents the ``conformal spin'' of the generator
$V^i_n$, while $-\infty <n< +\infty $ is the angular momentum
(the Fourier mode on the circle).
The first term on the right-hand-side of (\ref{Win}) reproduces the classical
$w_{\infty }$ algebra (\ref{clw}) by the correspondence
${\cal L}^{(cl)}_{i-n,i} \to V^i_n\ $ and identifies $W_{1+\infty}$ as
the algebra of ``quantum area-preserving diffeomorphisms''.
The additional terms are quantum operator corrections with polynomial
coefficients $q(i,j,n,m)$, due to the algebra of higher derivatives
\cite{shen}.
Moreover, the $c$-number term $c^i(n)$
is the quantum {\it anomaly}, a relativistic effect due to the
renormalization of operators acting on the infinite Dirac sea.
It is diagonal in the spin indices for our choice of basis for the
$\ g^i_k\ $ (see \cite{ctz4}).
Finally, the normal ordering ($\ :\ :\ $) of the Fock operators takes
care of the renormalization \cite{bpz} .

Let us analyse the generators $V^0_n$ and $V^1_n$ of lowest conformal
spin.
{}From (\ref{acf}) we see that the $V^0_n$ are Fourier modes of the fermion
density evaluated at the edge $\vert z\vert =R$;
thus, $V^0_0$ measures the edge charge.
Instead, the $V^1_n$ are vector fields which generate angular
momentum transformations on the edge, such that $V^1_0$ measures
the angular momentum of edge excitations.
Their algebra is given by
\beq
\left[\ V^0_n, V^0_m\ \right] = c\ n \ \delta_{n+m,0}\ \qquad\qquad,
\label{aff} \eeq
and
\barr
\left[\ V^1_n, V^0_m\ \right] & = & -m \ V^0_{n+m}\ ,\nonumber\\
\left[\ V^1_n, V^1_m\ \right] & = & (n-m) \ V^1_{n+m} + {c\over 12}\left(
n^3-n\right) \delta _{n+m,0}\ ,\label{Vir} \earr
with $c=1$.
These equations show that the $V^0_n$ and $V^1_n$ operators satisfy
the Abelian current (Kac-Moody) algebra $\u1$ and the Virasoro algebra,
respectively \cite{bpz}.
For unitary  $W_{1+\infty }$
theories, the central charge $c$ can be any
{\it positive integer} \cite{kac1} \cite{kac2} .

Following the standard procedure of two-dimensional
conformal field theory, we define a $W_{1+\infty}$ theory as the
Hilbert space given by a set
of irreducible, highest-weight representations of the $W_{1+\infty}$
algebra, closed under the {\it fusion rules} for making composite
excitations \cite{bpz} .
Any representation contains an infinite number of states, corresponding
to all the particle-hole excitations above a bottom state, the so-called
{\it highest weight} state.
This can be, for example, the ground-state $\vert\Omega\rangle$
corresponding to the incompressible quantum fluid.
The particle-hole excitations can be written as
\beq
V^{i_1}_{-n_1}\ V^{i_2}_{-n_2} \cdots V^{i_s}_{-n_s }
\vert\ \Omega\ \rangle\ ,\quad
n_1 \ge n_2\ge \cdots \ge n_s > 0\ ,
\quad i_1,\dots,i_s \ge 0 \ ,\label{neut} \eeq
while the positive modes $(n_i<0)$ annihilate $\vert\ \Omega\ \rangle$.
Here $k=\sum _j n_j$
is the total angular momentum of the edge excitation.
The number of independent states at fixed $k$, i.e.
the number of independent $V^i_n$ current modes, is finite; its actual
value depends on the
type of representation and the central charge.

Furthermore, any charged edge excitation, together with its tower of
particle-hole excitations, also forms an irreducible, highest-weight
representation of $W_{1+\infty}$. The states
in this representation have the same form of (\ref{neut}), but
the bottom state $\vert Q\rangle$ now represents a quasi-particle inside
the droplet.
The charge $Q$ and the spin $J$ of the quasi-particle
are given by the eigenvalues of the operators\footnote{
More precisely, $V^0_0$ measures the charge spilled to the edge, which
is minus the charge of the quasi-particle in the bulk, due to overall
charge conservation.} $V^0_0$ and $V^1_0$,
respectively:
\beq
V^0_0 \vert Q\rangle =Q\vert Q\rangle\ , \
V^1_0\vert Q\rangle = J \vert Q\rangle\ .\label{hws} \eeq
Moreover, the statistics $\theta/\pi$ of quasi-particles is equal to
twice the spin $J$.
For $\winf$ \reps, all the operators $V^i_0$  are simultaneously
diagonal and assign other quantum numbers to the quasi-particle,
$\ V^i_0 \vert Q\rangle =m_i(Q)\vert Q\rangle, \ i\ge 2\ $,
which are known polynomials in the charge $Q$ \cite{kac1}.
These quantum numbers measure the radial moments of the charge
distribution of a quasi-particle (see \cite{ctz4});
their fixed functional form indicates the rigidity of density
modulations of the quantum incompressible fluid.

\bigskip
\noindent{\bf Classification of QHE universality classes}

Besides the explicit example for $\nu=1$, leading to a theory
with $c=1$, it has been shown in general that the algebra (\ref{Win})
is the unique quantization of the $w_\infty$ algebra in the
$(1+1)$-dimensional field theory on the circle \cite{radul} .
We shall therefore characterize quantum incompressible fluids as
$W_{1+\infty }$ theories \cite{ctz3}.

This characterization provides a powerful classification scheme for
quantum Hall universality classes. These can in fact be classified by
using the recently developed representation theory \cite{kac1} \cite{kac2}
of $W_{1+\infty }$.
We shall classify quantum Hall universality classes by the following
{\it kinematical data}:

i) the quantum numbers $V^i_0$ of quasi-particle excitations, in particular
their charge and fractional statistics;

ii) the number of particle-hole excitations of given angular momentum,
i.e. the {\it degeneracies} of states on top of the ground state (\ref{neut});

iii) the Hall conductivity $\sigma_H=(e^2/h)\nu$, which is proportional
to the filling fraction $\nu$ of the ground state.

The Hall current produced by an external electric field
is actually given by the {\it chiral anomaly}
of the (1+1)-dimensional edge theory \cite{cdtz1}.


\bigskip
\section{Existing theories of edge excitations and experiments}

Before developing this classification program, we would like to briefly
review the existing theories of the quantum Hall effect and discuss
their description of the experimental data.

\vbox to 1. in {\vfill}
\noindent{\bf Hierarchical trial wave functions}

The Laughlin theory of the incompressible fluid \cite{laugh} was originally
developed for the Hall conductivities
$\ \sigma_{xy}=(e^2/h) \nu\ $, where $\ \nu=1,1/3,1/5,1/7,\dots$
are the filling fractions.
Afterwards, a hierarchical generalization of these trial wave functions
was introduced by Haldane and Halperin \cite{hald}, in order to
describe other observed filling fractions.
Therefore, by the {\it hierarchy} problem we usually mean the classification
of stable ground states (and their excitations) corresponding
to all observed plateaus.
Naturally, the stability is related to the order of iteration
of the hierarchical construction, starting from the integer fillings,
then the Laughlin fillings and so forth.
The Haldane-Halperin hierarchy is not completely satisfactory, because
it produces ground states for too many filling fractions, already
at low order of iteration.
On the contrary, the experiments show only some stable ground states
(see fig. 1).
Although numerical experiments show that the hierarchical wave
functions are rather accurate, their construction lacks a good control
of stability.

Another hierarchical construction of wave functions, which match most of
the experimental plateaus to lowest order of the hierarchy, has been
proposed by Jain \cite{jain}.
Jain abstracted from Laughlin's work the concept of {\it composite
fermion}, a local bound state of the electron and an even number of
flux quanta.
Due to yet unknown dynamical reasons, the composite fermions are
stable quasi-particles, which interact weakly among themselves.
Moreover, the strongly-interacting electrons at fractional filling can be
mapped into composite fermions at effective integer filling.
Therefore, the stability of the observed ground states
with fractional filling can be related to the stability of completely
filled Landau levels.
The composite fermion picture was successfully applied \cite{hlr} to the
independent dynamics of the compressible fluid at $\nu=1/2$.
This strongly-interacting, gapless ground state can be described
as a Fermi liquid of composite fermions with vanishing effective
magnetic field.
Experiments \cite{nu1/2} have confirmed this theory by observing the free
motion of the composite fermions.

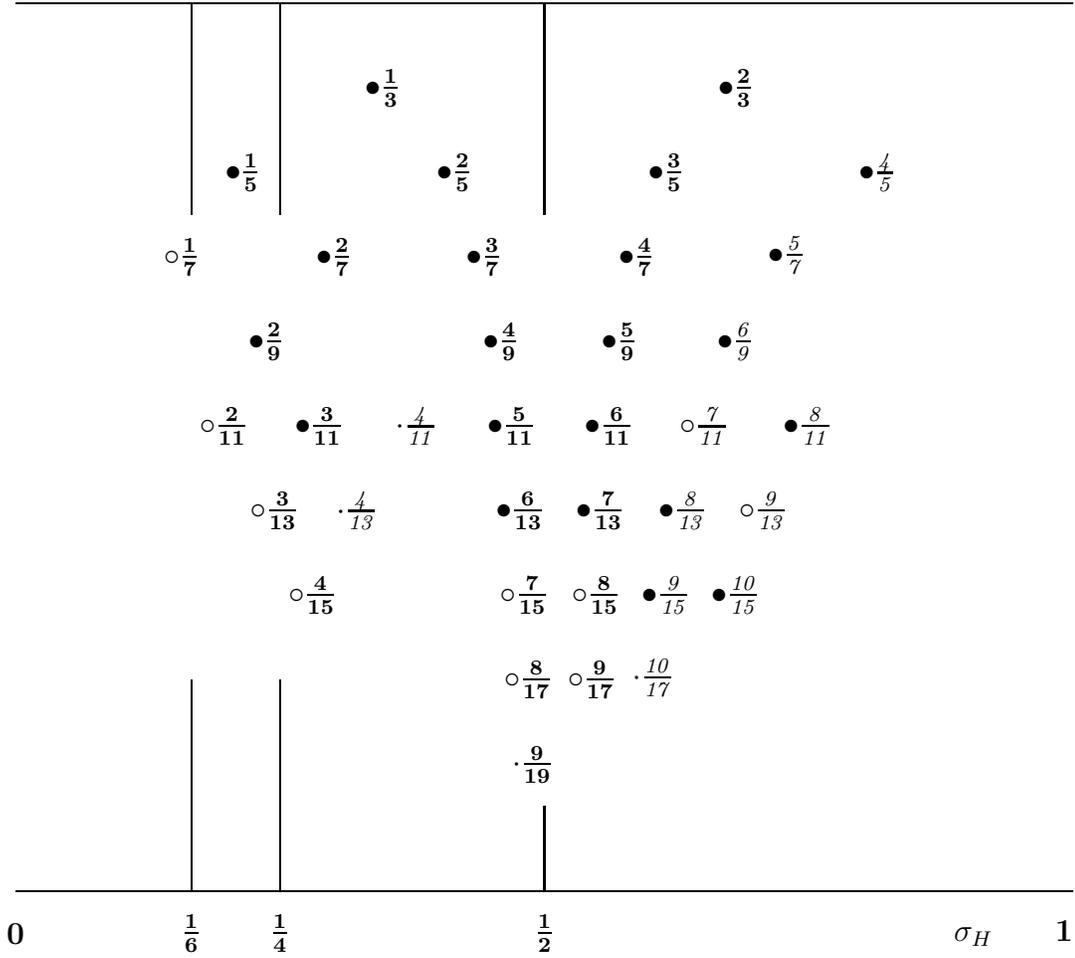
\begin{figure}
\unitlength=0.8pt
\begin{picture}(500.00,440.00)(-10.00,0.00)
\put(0.00,500.00){\line(1,0){500.00}}
\put(0.00,80.00){\line(1,0){500.00}}
\put(250.00,500.00){\line(0,-1){100.00}}
\put(125.00,500.00){\line(0,-1){100.00}}
\put(83.00,500.00){\line(0,-1){100.00}}
\put(250.00,80.00){\line(0,1){40.00}}
\put(125.00,80.00){\line(0,1){100.00}}
\put(83.00,80.00){\line(0,1){100.00}}
\put(166.00,460.00){\makebox(0,0)[cc]{$\ \ \ \bullet {\bf 1\over 3}$}}
\put(333.00,460.00){\makebox(0,0)[cc]{$\ \ \ \bullet {\bf 2\over 3}$}}
\put(100.00,420.00){\makebox(0,0)[cc]{$\ \ \ \bullet {\bf 1\over 5}$}}
\put(200.00,420.00){\makebox(0,0)[cc]{$\ \ \ \bullet {\bf 2\over 5}$}}
\put(300.00,420.00){\makebox(0,0)[cc]{$\ \ \ \bullet {\bf 3\over 5}$}}
\put(400.00,420.00){\makebox(0,0)[cc]{$\ \ \ \bullet {\it4\over 5}$}}
\put(71.00,380.00){\makebox(0,0)[cc]{$\ \ \ \circ    {\bf 1\over 7}$}}
\put(143.00,380.00){\makebox(0,0)[cc]{$\ \ \ \bullet {\bf 2\over 7}$}}
\put(214.00,380.00){\makebox(0,0)[cc]{$\ \ \ \bullet {\bf 3\over 7}$}}
\put(286.00,380.00){\makebox(0,0)[cc]{$\ \ \ \bullet {\bf 4\over 7}$}}
\put(357.00,380.00){\makebox(0,0)[cc]{$\ \ \ \bullet {\it 5\over 7}$}}
\put(111.00,340.00){\makebox(0,0)[cc]{$\ \ \ \bullet {\bf 2\over 9}$}}
\put(222.00,340.00){\makebox(0,0)[cc]{$\ \ \ \bullet {\bf 4\over 9}$}}
\put(278.00,340.00){\makebox(0,0)[cc]{$\ \ \ \bullet {\bf 5\over 9}$}}
\put(333.00,340.00){\makebox(0,0)[cc]{$\ \ \ \bullet {\it 6\over 9}$}}
\put(91.00,300.00){\makebox(0,0)[cc]{$\ \ \ \circ    {\bf 2\over 11}$}}
\put(136.00,300.00){\makebox(0,0)[cc]{$\ \ \ \bullet {\bf 3\over 11}$}}
\put(182.00,300.00){\makebox(0,0)[cc]{$\ \ \ \cdot   {\it 4\over 11}$}}
\put(227.00,300.00){\makebox(0,0)[cc]{$\ \ \ \bullet {\bf 5\over 11}$}}
\put(273.00,300.00){\makebox(0,0)[cc]{$\ \ \ \bullet {\bf 6\over 11}$}}
\put(318.00,300.00){\makebox(0,0)[cc]{$\ \ \ \circ   {\it7\over 11}$}}
\put(367.00,300.00){\makebox(0,0)[cc]{$\ \ \ \bullet {\it8\over 11}$}}
\put(115.00,260.00){\makebox(0,0)[cc]{$\ \ \ \circ   {\bf 3\over 13}$}}
\put(154.00,260.00){\makebox(0,0)[cc]{$\ \ \ \cdot   {\it4\over 13}$}}
\put(231.00,260.00){\makebox(0,0)[cc]{$\ \ \ \bullet {\bf 6\over 13}$}}
\put(269.00,260.00){\makebox(0,0)[cc]{$\ \ \ \bullet {\bf 7\over 13}$}}
\put(308.00,260.00){\makebox(0,0)[cc]{$\ \ \ \bullet {\it8\over 13}$}}
\put(346.00,260.00){\makebox(0,0)[cc]{$\ \ \ \circ   {\it9\over 13}$}}
\put(133.00,220.00){\makebox(0,0)[cc]{$\ \ \ \circ   {\bf 4\over 15}$}}
\put(233.00,220.00){\makebox(0,0)[cc]{$\ \ \ \circ   {\bf 7\over 15}$}}
\put(267.00,220.00){\makebox(0,0)[cc]{$\ \ \ \circ   {\bf 8\over 15}$}}
\put(300.00,220.00){\makebox(0,0)[cc]{$\ \ \ \bullet {\it9\over 15}$}}
\put(333.00,220.00){\makebox(0,0)[cc]{$\ \ \ \bullet {\it 10\over 15}$}}
\put(235.00,180.00){\makebox(0,0)[cc]{$\ \ \ \circ   {\bf 8\over 17}$}}
\put(265.00,180.00){\makebox(0,0)[cc]{$\ \ \ \circ   {\bf 9\over 17}$}}
\put(294.00,180.00){\makebox(0,0)[cc]{$\ \ \ \cdot   {\it10\over 17}$}}
\put(237.00,140.00){\makebox(0,0)[cc]{$\ \ \ \cdot   {\bf 9\over 19}$}}
\put(83.00,60.00){\makebox(0,0)[cc]{${\bf 1\over 6}$}}
\put(125.00,60.00){\makebox(0,0)[cc]{${\bf 1\over 4}$}}
\put(250.00,60.00){\makebox(0,0)[cc]{${\bf 1\over 2}$}}
\put(0.00,60.00){\makebox(0,0)[cc]{${\bf 0}$}}
\put(500.00,60.00){\makebox(0,0)[rc]{$\sigma_H \qquad {\bf 1}$}}
\end{picture}
\caption{
Experimentally observed plateaus in the range $0<\nu<1$ : their Hall
conductivity $\sigma_H=(e^2/h)\nu$ is displayed in units of $(e^2/h)$.
The points denote stability:
$\ (\bullet)\ $ very stable, $\ (\circ)\ $ stable, and
$\ (\cdot)\ $ less stable plateaus. Theoretically understood plateaus are in
{\bf bold}, unexplained ones are in {\it italic}.
Observed cases of coexisting fluids are displayed
as $\nu=2/3,6/9,10/15$,  $\ \nu=3/5,9/15$ and $\nu=5/7, 15/21$ (but
$15/21$ is not displayed). (Adapted from ref. [31])}
\end{figure}

\bigskip
\noindent{\bf The chiral boson theory of the edge excitations}

After the original works of Halperin \cite{halp} and Stone \cite{stone},
a general theory of edge excitations, corresponding to the hierarchical
constructions of wave-functions, has been formulated
\cite{jtrans}\cite{wen}\cite{kmat}.
This is the $(1+1)$-dimensional theory of the chiral boson \cite{flo}.
An equivalent description is given by
Abelian Chern-Simons theories on $(2+1)$-dimensional open domains \cite{kmat}.
The edge excitations of the Laughlin fluid are described by a one-component
chiral boson, while the hierarchical fluids require many components.
Every boson describes an independent edge current, and thus the
incompressible fluids have generically a composite edge structure.
Each current gives rise to the Abelian current algebra (\ref{aff}),
 which implies the Virasoro algebra (\ref{Vir}) with
central charge $c=1$ \cite{bpz}.

On an annulus geometry, with edge circles $|{\bf x}|=R_1$ and $|{\bf x}|=R_2$,
one introduces $m$ independent one-dimensional {\it chiral} currents
\beq
J^i \left(R_1\theta-v_i t\right) = - {1\over 2\pi R_1}
{\partial\over\partial\theta} \ \phi^i \ , \qquad (|{\bf x}|=R_1),
\label{jchi}\eeq
and corresponding ones with opposite chirality
$J^i \left(R_2\theta+v_i t\right)$ at the other edge $|{\bf x}|=R_2$.
The dynamics of these currents on the edge circle $|{\rm x}|=R_1$
is governed by the action,
\beq
S=-{1\over 4\pi}\ \int\ dt\ dx\ \sum_{i=1}^m\ \kappa_i
\left(\partial_t\phi^i +v_i \partial_x\phi^i \right)
\partial_x\phi^i\ , \qquad {\rm for}\ \ \  x\equiv R_1\theta\ ,
\label{bosact}\eeq
for the $m$ $(1+1)$-dimensional {\it chiral boson} fields $\phi^i$ \cite{flo}.
The corresponding action for the other circle
$x\equiv R_2\theta$ is obtained by replacing $v_i\to (-v_i)$.
The dynamics on the two edges are identical and independent, only constrained
by the conservation of the total charge: thus we describe one of them only.
We can change the normalization of the fields, and reduce each
coupling constant to a sign, $\kappa_i \to \pm 1$.
The equations of motion imply that the fields are
chiral, $\phi^i=\phi^i(x-v_it)\ $, and canonical quantization
implies the following commutation relations for the currents,
\beq
{[ J^i(x_1),J^k(x_2) ]}= {1\over 2\pi\kappa_i}\ \delta^{ik}\
\delta^{\prime}(x_1-x_2)\ ,\qquad (t_1=t_2) \ ,
\label{curalg}\eeq
which are those of the multi-component Abelian current algebra
$\widehat{U(1)}^{\otimes m}$ \cite{bpz}.
The positive definiteness of the Hamiltonian requires the signs of the
velocities $v_i$ and the couplings $\kappa_i$ to be related:
$v_i\kappa_i > 0\ , \qquad i=1,\dots ,m\ $.

Let us discuss one particular chiral current, $v_i>0$
(i.e. $\kappa_i=1$). The quantization of the chiral boson is equivalent to
the construction of the representations of
the current algebra (\ref{curalg}). Actually, all the states in the
Hilbert space of the theory can be fitted into a set of
representations \cite{bpz}.
To this end, we introduce the Fourier modes of the currents,
\beq
J^i(R\theta-v_it)={1\over 2\pi}\ \sum_{n=-\infty}^{\infty}\
\alpha^i_n\ {\rm e}^{in(\theta-v_it)}\ ,
\label{mod1}\eeq
which satisfy,
\beq
{[}\alpha^i_n,\alpha^j_m {]}= \ \delta^{ij}\ {n\over \kappa_i}\
\delta_{n+m,0}\ .
\label{km}\eeq
The positivity of the ground-state expectation value
$\langle\Omega |\alpha^i_n \alpha^i_{-n} |\Omega\rangle
\equiv \Vert\alpha^i_n |\Omega\rangle\Vert^2\ \ge 0\ $,
and the commutation relations (\ref{km}) with $\kappa_i=1$ imply the conditions
\beq
\alpha^i_n |\Omega\rangle=0\ , \qquad n>0 \qquad\quad (v_i >0).
\label{hwsc}\eeq
An irreducible highest-weight representation of the
$\widehat{U(1)}$ current algebra is made by the highest-weight state
$|\Omega\rangle$ and by all states obtained by applying any
number of $\alpha^i_n, \ \ n<0\ ,$ to it.
The weight of the representation is given by the eigenvalue of $\alpha^i_0$,
which is the single-edge charge, in units to be specified later.
For the ground state, we have
\beq
\alpha^i_0 |\Omega\rangle = 0\ .
\label{cvac}\eeq
Other unitary highest-weight representations can be similarly built on top
of other highest-weight states $|r\rangle$, $r\in {\rm R}$, which satisfy
\beq
\alpha^i_n |r\rangle= 0 \quad n>0\ ,\qquad \alpha^i_0|r\rangle = r
|r\rangle\ ,
\label{hwsr}\eeq
and are built by applying the {\it vertex operators} to the
ground state \cite{bpz}.
These \reps correspond to the quasi-particle excitations of this edge theory.
The Virasoro generators are defined by the Sugawara construction \cite{bpz},
\beq
L^i_n ={\kappa_i\over 2}\ \sum_{l=-\infty}^{\infty}\ :\
\alpha^i_{n-l}\alpha^i_l\ :\ .
\label{ln}\eeq
They give rise to the Virasoro algebra (\ref{Vir}) with $c=1$, for each
current component $i$.

The $m$-edge theory has $\u1^{\otimes m}$ symmetry, $c=m$, and is parametrized
by an integer, symmetric $(m\times m)$ matrix, with odd diagonal elements,
the $K$ matrix \cite{kmat}, which
determines the Hall conductivity and the fractional charge, spin
and statistics of the edge excitations.
In the chiral boson theory, the fusion rules are the addition of weight
vectors $\vec{r}$; the set of \reps which is closed under these rules is
the {\it lattice} $\Gamma$,
\beq
\Gamma=\left\{ \left. \vec{r}\ \right\vert\ \vec{r}=
\sum_{i=1}^m n_i \vec{v}_i\ ,\quad n_i \in {\bf Z} \right\} \ .
\label{latt}\eeq
The basis vectors $\vec{v}_i$  represent a physical elementary excitation
in the $i$-th edge component, which may not correspond to the previous basis
of propagating modes. The {\it physical charge} of an
excitation with labels $n_i\in {\rm Z}$ is thus given by the sum of
the components in the physical basis \cite{ctz5},
\beq
Q  = \sum_{i,j=1}^m\ K^{-1}_{ij}\ n_j \ ,
\label{qform}\eeq
where
\beq
K^{-1}_{ij} =\sum_{l=1}^m\ \Lambda_{il} {1\over \kappa_l} \Lambda^T_{lj} =
\left( \vec{v}_i\cdot \eta\cdot \vec{v}_j \right) \ .
\label{met}\eeq
Similarly, the fractional spin and statistics of this excitation is given
by the sum of the eigenvalues of the Virasoro generators $L^i_0$,
\beq
{ \theta\over \pi} = \sum_{i,j=1}^m\ n_i \ K^{-1}_{ij}\ n_j \ ,
\qquad n_i \in {\bf Z} \ .
\label{qtheta}\eeq
In general, the metric $K^{-1}$ of the lattice $\Gamma$ in the basis
$\vec{v}_i$ is pseudo-Euclidean with signature
$\eta_{ij}=\delta_{ij}\kappa_i$, due to the possible presence of excitations
with different chiralities.

The Hall conductivity in the annulus geometry can be measured by applying
a uniform electric field along all the edges, $E^i=E$.
The chiral anomaly of the edge theory actually corresponds to a radial
flow of particles in the annulus, which move from the inner edge
to the outer edge. The Hall conductivity can be thus found to be \cite{cdtz1},
\beq
\sigma_H={e^2\over h}\ \nu \ ,\qquad \nu= \sum_{i,j=1}^m\ K^{-1}_{ij} \ .
\label{sig}\eeq
Equations (\ref{qform}-\ref{sig}) for the Hall conductivity and
the spectrum of the charge and fractional statistics of edge excitations
are the basic data of the quantum incompressible fluid described by
the chiral boson theories \cite{wen}\cite{kmat}.
The existence of $m$ electron excitations with unit charge
and integer statistic relative to all excitations,  requires that
$K$ has integer entries with odd integers on the diagonal \cite{kmat}.

\bigskip

\noindent{\bf The Jain hierarchy}

The Jain fluids have been described by the subset of the chiral boson theories
characterized by the following $K$ matrices \cite{kmat},
\beq
K_{ij}=\pm\delta_{ij} + p\ C_{ij}\ , \qquad C_{ij}=1\ \forall\ i,j=1,\dots,m\ ,
\ p>0\ {\rm even}\ ,
\label{kjain}\eeq
and the following spectra of edge excitations,
\barr
\nu &= &{m\over mp \pm 1}\ , \qquad p >0 \ {\rm even}\ ,\qquad (c =m)\ ,
\nonumber\\
Q & = & {1\over pm \pm 1}\ \sum_{i=1}^m n_i \ ,\nonumber\\
{ \theta\over\pi} &=& \pm\left( \sum_{i=1}^m n^2_i -
{p\over mp \pm 1}\left( \sum_{i=1}^m n_i \right)^2 \right) \ .
\label{jspec}\earr
Note that $K$ has $(m-1)$ degenerate eigenvalues $\lambda_i =1$,
$\ i=1,\dots, m-1\ $ (resp. $\lambda_i=-1$),
and a single value $\lambda_m=\pm 1+mp$. If the sign $\pm$ is negative,
one edge has opposite chirality to the others.
There is one basic charged quasi-particle excitation with label
$n_i=(1,\dots,1)$ and $m(m-1)/2$ {\it neutral} excitations for
$n_i=(\delta_{ik}-\delta_{il})$, $\ 1\le k <l\le m$, with identical integer
statistics.

The corresponding trial wave functions for the ground state
have been constructed by Jain as \cite{jain}
\beq
\Psi_{\nu}=D^{p/2}\ L^m\ {\bf 1}\ , \qquad p\ {\rm even} \ ,
\label{jjain}\eeq
where $L^m\ {\bf 1}$ represents schematically the wave function of $m$ filled
Landau levels and $D^{p/2}$ multiplies the wave
function by the $p$-th power of the Vandermonde determinant, which
``attaches $p$ flux tubes to each electron'', and transforms them
into ``composite fermions''.

The Jain hierarchy covers most of the experimentally observed plateaus,
as we discuss in the next section.
However, within the chiral boson approach, there is no clear motivation for
selecting the special $K$ matrices (\ref{kjain}).
The size of the gap for bulk density waves is usually invoked for
solving this puzzle: the observed fluids are supposed
to have the largest gaps, while the general $K$ fluids have small gaps
and are destroyed by thermal fluctuations and other effects.
It is also true that edge theories give kinematically possible incompressible
fluids and their universal properties, but cannot describe
the size of the gaps, which is determined by the microscopic
bulk dynamics.

Nevertheless, we have found a natural way to select
the Jain hierarchy within the $\winf$ edge theory approach \cite{ctz5}.
Indeed, the Jain fluids correspond to the $\winf$ minimal models, which are
characterized by possessing less states than their chiral boson counterparts.
We propose this reduction of available states as a natural stability
principle.

\bigskip

\noindent{\bf Experiments}

We first discuss the spectrum of fractional Hall conductivities in
eq. (\ref{jspec}). According to Jain, the stability of the ground states
(\ref{jjain}) should be approximately independent of $m$, which counts the
number of Landau levels filled by the composite fermion.
This is in analogy with the integer Hall plateaus, which are all
equally stable. On the other hand, the stability decreases
by increasing $|p|$, as observed for the Laughlin fluids ($m=1$).
Therefore, the most stable family of plateaus is,
\beq
\nu={m\over 2m \pm 1} \ ,\qquad (p=2)\ ,
\label{fam1}\eeq
which accumulate at $\nu=1/2$. The next less stable family is
\beq
\nu={m\over 4m \pm 1} \ ,\qquad (p=4)\ ,
\label{fam2}\eeq
which accumulate at $\nu=1/4$.
This behaviour is clearly seen in the experimental data of fig. 1.
For these filling fractions, the Jain wave functions for the ground state
and the simplest excited states have a good overlap with those obtained
numerically by diagonalizing the microscopic Hamiltonian
with a small number of electrons.

A closer look into the experimental values of the filling fractions
in fig. 1 shows other points (in italic), like $\nu=4/5,\ 5/7,\ 8/11$,
which fall outside the  Jain main series (\ref{fam1},\ref{fam2}) (in bold).
These points were originally interpreted as ``charge conjugates'' of these
series \cite{jain},
\beq
\nu=1-{m\over 2m \pm 1} \ , \qquad \nu=1 -{m\over 4m \pm 1} \ ,
\label{ccon}\eeq
which actually belong to the second
iteration of the Jain hierarchy \cite{jain}.
Unfortunately, the charge conjugate states do not fit well the data in
fig. 1. The $\nu=1/2$ family would be self-conjugate; thus there should
be two fluids per filling fraction, which are not observed, apart from two
cases.
Actually, coexisting fluids can be detected by experiments where the magnetic
field is tilted from the orthogonal direction to the plane \cite{tilt}.
Furthermore, the conjugate of the observed fractions in the
$\nu=1/4$ family are not observed in half of the cases.
Finally, there are fractions which do not belong to any previous group:
$\nu=4/11,\ 7/11,\ 4/13,\ 8/13$, $\ 9/13,\ 10/17$.

In conclusion, all the fractions outside the main Jain families
((\ref{fam1},(\ref{fam2}) are not well understood at present (and will not be
explained in this paper).
Any known extension of the previous theory which explains these few extra
fractions, also introduces many more unobserved fractions, with an unclear
pattern of stability.
Besides the second iteration of the Jain hierarchy \cite{jain}, we also quote
the approach proposed by Fr\"ohlich and collaborators \cite{froh}.
They analyzed all lattices $\Gamma$ (\ref{latt}), with positive-definite,
integer (inverse) metric $K$, for small values of $det(K)$, whose
classification is known in the mathematical literature. These lattices can
be related to the $SU(m)$, $SO(k)$ and exceptional Lie algebras.
The stability of the corresponding fluids does not follow a clear pattern
related to these algebras, besides the case of the chiral Jain fluids
(\ref{kjain}), whose $SU(m)$ symmetry \cite{read} will explained in
the next section.
Moreover, in this approach, the $K$ matrices for the Jain filling fractions
$\nu=m/(mp-1)\ ,p>0$, are different from the Jain proposal (\ref{kjain})
which is not positive definite.

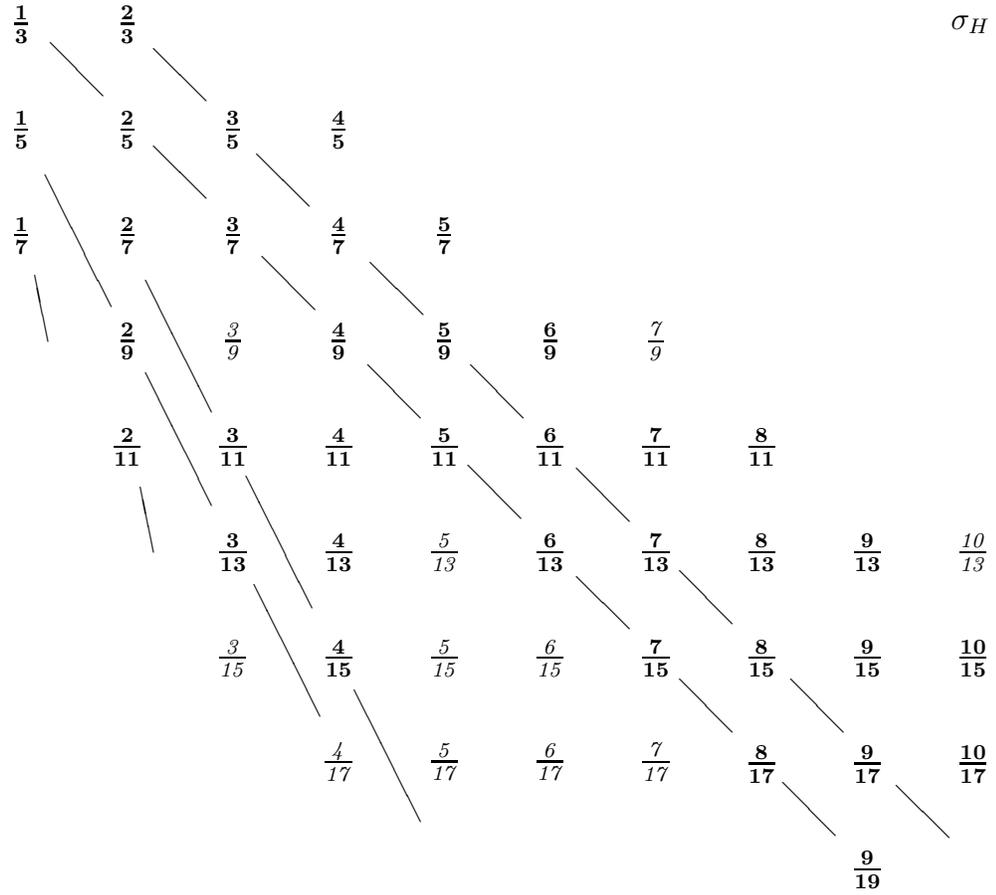
\begin{figure}
\unitlength=1.0pt
\begin{picture}(440.00,440.00)(0.00,0.00)
\put(10.00,500.00){\line(1,0){420.00}}
\put(10.00,100.00){\line(1,0){420.00}}
\put(348.00,153.00){\line(-1,1){20.00}}
\put(309.00,192.00){\line(-1,1){20.00}}
\put(270.00,231.00){\line(-1,1){20.00}}
\put(229.00,273.00){\line(-1,1){20.00}}
\put(191.00,311.00){\line(-1,1){20.00}}
\put(151.00,352.00){\line(-1,1){20.00}}
\put(110.00,394.00){\line(-1,1){20.00}}
\put(71.00,433.00){\line(-1,1){20.00}}
\put(351.00,192.00){\line(-1,1){20.00}}
\put(309.00,233.00){\line(-1,1){20.00}}
\put(270.00,272.00){\line(-1,1){20.00}}
\put(230.00,311.00){\line(-1,1){20.00}}
\put(192.00,350.00){\line(-1,1){20.00}}
\put(149.00,391.00){\line(-1,1){20.00}}
\put(110.00,431.00){\line(-1,1){20.00}}
\put(391.00,152.00){\line(-1,1){20.00}}
\put(153.00,198.00){\line(-1,2){25.00}}
\put(112.00,278.00){\line(-1,2){25.00}}
\put(74.00,353.00){\line(-1,2){25.00}}
\put(112.00,313.00){\line(-1,2){25.00}}
\put(150.00,239.00){\line(-1,2){25.00}}
\put(191.00,158.00){\line(-1,2){25.00}}
\put(50.00,340.00){\line(-1,5){5.00}}
\put(90.00,260.00){\line(-1,5){5.00}}
\put(40.00,460.00){\makebox(0,0)[cc]{$ {\bf 1\over 3}$}}
\put(80.00,460.00){\makebox(0,0)[cc]{$ {\bf 2\over 3}$}}
\put(40.00,420.00){\makebox(0,0)[cc]{$ {\bf 1\over 5}$}}
\put(80.00,420.00){\makebox(0,0)[cc]{$ {\bf 2\over 5}$}}
\put(120.00,420.00){\makebox(0,0)[cc]{$ {\bf 3\over 5}$}}
\put(160.00,420.00){\makebox(0,0)[cc]{$ {\bf 4\over 5}$}}
\put(40.00,380.00){\makebox(0,0)[cc]{$  {\bf 1\over 7}$}}
\put(80.00,380.00){\makebox(0,0)[cc]{$ {\bf 2\over 7}$}}
\put(120.00,380.00){\makebox(0,0)[cc]{$ {\bf 3\over 7}$}}
\put(160.00,380.00){\makebox(0,0)[cc]{$ {\bf 4\over 7}$}}
\put(200.00,380.00){\makebox(0,0)[cc]{$ {\bf 5\over 7}$}}
\put(80.00,340.00){\makebox(0,0)[cc]{$ {\bf 2\over 9}$}}
\put(120.00,340.00){\makebox(0,0)[cc]{$ {\it  3\over 9}$}}
\put(160.00,340.00){\makebox(0,0)[cc]{$ {\bf 4\over 9}$}}
\put(200.00,340.00){\makebox(0,0)[cc]{$ {\bf 5\over 9}$}}
\put(240.00,340.00){\makebox(0,0)[cc]{$ {\bf 6\over 9}$}}
\put(280.00,340.00){\makebox(0,0)[cc]{$ {\it  7\over 9}$}}
\put(80.00,300.00){\makebox(0,0)[cc]{$  {\bf 2\over 11}$}}
\put(120.00,300.00){\makebox(0,0)[cc]{$ {\bf 3\over 11}$}}
\put(160.00,300.00){\makebox(0,0)[cc]{$ {\bf 4\over 11}$}}
\put(200.00,300.00){\makebox(0,0)[cc]{$ {\bf 5\over 11}$}}
\put(240.00,300.00){\makebox(0,0)[cc]{$ {\bf 6\over 11}$}}
\put(280.00,300.00){\makebox(0,0)[cc]{$ {\bf 7\over 11}$}}
\put(320.00,300.00){\makebox(0,0)[cc]{$ {\bf 8\over 11}$}}
\put(120.00,260.00){\makebox(0,0)[cc]{$ {\bf 3\over 13}$}}
\put(160.00,260.00){\makebox(0,0)[cc]{$ {\bf 4\over 13}$}}
\put(200.00,260.00){\makebox(0,0)[cc]{$ {\it  5\over 13}$}}
\put(240.00,260.00){\makebox(0,0)[cc]{$ {\bf 6\over 13}$}}
\put(280.00,260.00){\makebox(0,0)[cc]{$ {\bf 7\over 13}$}}
\put(320.00,260.00){\makebox(0,0)[cc]{$ {\bf 8\over 13}$}}
\put(360.00,260.00){\makebox(0,0)[cc]{$ {\bf 9\over 13}$}}
\put(400.00,260.00){\makebox(0,0)[cc]{$ {\it  10\over 13}$}}
\put(120.00,220.00){\makebox(0,0)[cc]{$ {\it  3\over 15}$}}
\put(160.00,220.00){\makebox(0,0)[cc]{$ {\bf 4\over 15}$}}
\put(200.00,220.00){\makebox(0,0)[cc]{$ {\it  5\over 15}$}}
\put(240.00,220.00){\makebox(0,0)[cc]{$ {\it  6\over 15}$}}
\put(280.00,220.00){\makebox(0,0)[cc]{$ {\bf 7\over 15}$}}
\put(320.00,220.00){\makebox(0,0)[cc]{$ {\bf 8\over 15}$}}
\put(360.00,220.00){\makebox(0,0)[cc]{$ {\bf 9\over 15}$}}
\put(400.00,220.00){\makebox(0,0)[cc]{$ {\bf 10\over 15}$}}
\put(160.00,180.00){\makebox(0,0)[cc]{$ {\it  4\over 17}$}}
\put(200.00,180.00){\makebox(0,0)[cc]{$ {\it  5\over 17}$}}
\put(240.00,180.00){\makebox(0,0)[cc]{$ {\it  6\over 17}$}}
\put(280.00,180.00){\makebox(0,0)[cc]{$ {\it  7\over 17}$}}
\put(320.00,180.00){\makebox(0,0)[cc]{$ {\bf 8\over 17}$}}
\put(360.00,180.00){\makebox(0,0)[cc]{$ {\bf 9\over 17}$}}
\put(400.00,180.00){\makebox(0,0)[cc]{$ {\bf 10\over 17}$}}
\put(360.00,140.00){\makebox(0,0)[cc]{$ {\bf 9\over 19}$}}
\put(430.00,460.00){\makebox(0,0)[rc]{$\sigma_H \qquad $}}
\end{picture}
\caption{
List of all fractions $\nu=p/q$, with $2/11<\nu<4/5$, $1<p\le 10$ and
$3\le q\le 17$, $\ q$ odd.
The fractions corresponding to experimental values of the Hall
conductivity $\sigma_H=(e^2/h)\nu$ are in {\bf bold};
the unobserved fractions
are in {\it italic}. Observed fractions joined by lines
are explained by the Jain hierarchy (3.18,3.19).}
\end{figure}

In figure 2, we study the limitations of phenomenological descriptions
of the stability of the fluids. Besides all the observed (bold)
fractions of fig. 1, we report the unobserved (italic) ones $\nu=p/q$,
which satisfy the conservative cuts of the ``phase space''
\beq
{2\over 11} < \nu={p\over q} < {4\over 5} \ , \qquad {\rm and}\ \
p \le 10\ , q\le 17 \ .
\label{phasp}\eeq
Namely, we display all fractions which would be observed if
the gap were a smooth function of the parameters $(\nu,p,q)$ interpolating
the data, a typical phenomenological hypothesis.
Figure 2 shows that, besides the families ((\ref{fam1},(\ref{fam2}), about
half of the fractions are unexplained observed fillings
and half are unobserved but phenomenologically possible.
This implies that the gap is not a smooth function of simple parameters
like $(\nu,p,q)$ - deeper theories are needed to explain stability.

Actually, a major virtue of the Jain hierarchy is that of representing
one-parameter families of Hall states, within which the gap {\it is}
a smooth function of the above parameters.
We think of these families as the set of {\it kinematically allowed}
quantum incompressible fluids (at first level of the hierarchy).

More specific confirmations of the edge theory (\ref{kjain}) come
from the experimental tests of the spectrum of excitations (\ref{jspec}).
An experiment with high time resolution \cite{tdom} has measured the
propagation
of a single chiral charge excitation for $\nu=1/3\ $ ($m=1\ ,p=2$),
and $\nu=2/3\ $ ($m=2,\ p=2$); this is in agreement with the Jain theory,
although the neutral excitations have not been seen yet.
The resonant tunnelling experiment \cite{milli} has verified the conformal
dimensions (\ref{jspec}) for the simplest Laughlin fluid $\nu=1/3$ \cite{tunn}.
Extensions of this experiment to $(m>1)$ fluids have been suggested, as well
as tests of the neutral edge spectrum \cite{kane}.
We shall discuss more these experiments in section 4.


\section{$\winf$ minimal models}

\noindent{\bf The theory of $\winf$ \reps }

We now develop the classification program of incompressible quantum Hall
fluids outlined in section two. The basic piece of information we need
is the mathematical theory of $\winf$ \reps. Luckily enough,
all unitary, irreducible (quasi-finite) $\winf$ representations
were obtained in the
fundamental work by Kac and collaborators \cite{kac1}\cite{kac2}:
they exist for integer central charge $(c=m)$ and can be regular, i.e.
{\it generic}, or {\it degenerate}.
In ref.\cite{ctz3}, we used the generic representations to build the
{\it generic} $\winf$ theories, which were shown to
correspond to the previously described $m$-component chiral boson
theories parametrized by generic $(m\times m)$ $K$ matrices.
In the algebraic approach, this is proven by identifying
the generic $\winf$ \reps with $\u1^{\otimes m}$ \reps.
Both \reps are labelled by the same weight vectors
$\vec{r}$. A complete equivalence requires also a one-to-one map between
the states built on top of the respective highest weight states.
The general theory of $\winf$ \reps \cite{kac1}\cite{kac2}, leads to the
following relations between irreducible \reps of the two algebras,
\beq\begin{array}{l l l}
M\left(\winf ,1,r\right) &\sim& M\left(\widehat{U(1)},1,r\right)\ ;\\
M\left(\winf ,m>1,\vec{r}\right) &\sim&
M\left(\widehat{U(1)}^{\otimes m},m,\vec{r}\right)\ ,
\qquad {\rm for}\ (r_i-r_j) \not\in {\bf Z} \ ,\forall\ i\neq j\ , \\
M\left(\winf ,m>1,\vec{r}\right) & \subset &
M\left(\widehat{U(1)}^{\otimes m},m,\vec{r} \right)\ ,
\qquad {\rm if \ }\exists\ (r_i - r_j) \in {\bf Z} \ .
\end{array}\label{winclu}\eeq
Generically, $\winf$ and $\u1^{\otimes m}$ \reps are one-to-one equivalent.
The exceptions appear for $c>1$, when the weight has some integer
components $(r_i - r_j)$. In these cases, the relation is many-to-one, i.e. an
irreducible $\widehat{U(1)}^{\otimes m}$ representation is {\it reducible}
with respect to the $\winf$ algebra.
We call {\it generic} the $\winf$ \reps which are one-to-one equivalent to
$\u1^{\otimes m}$ ones ($(r_i-r_j) \not\in {\bf Z} \ ,\forall\ i\neq j$),
and {\it degenerate} the remaining $\winf$ \reps
($\exists\ (r_i - r_j) \in {\bf Z}$).

The results (\ref{winclu}) allow the construction of several
types of $\winf$ symmetric theories.
The generic $\winf$ theories \cite{ctz3} are defined by lattices $\Gamma$
(\ref{latt}) which contain generic $\winf$ \reps only:
for these, the basis vectors satisfy
\hbox{ $((\vec{v}_\alpha)_i -(\vec{v}_\alpha)_j)\not\in {\bf Q}$},
$\forall\ \alpha, i\neq j =1,\dots, m$.
These theories are thus equivalent to chiral boson theories\footnote{
The ground state representation ($\vec{r}=0$) must also be a $\u1^{\otimes m}$
\rep for the closure of the fusion rules.}.
Other $\winf$ theories, containing only degenerate \reps, are instead
different. These are the {\it minimal models}, which we shall describe
below. They are the basic new $\winf$ theories,
and are actually very important, because they will be shown to
correspond to the experimentally observed Jain fluids.
The mathematical rules for building the degenerate representations
have a hierarchical structure similar to the Jain construction:
in ref. \cite{ctz5}, we fully explained this correspondence to the lowest
order of the hierarchies.

On the other hand, the chiral boson theories of the Jain hierarchy
(\ref{jspec})
have been widely used in the literature and partially confirmed by
the experiments, as we discuss hereafter.
These are also $\winf$ symmetric, but are not the simplest realizations of
this symmetry, because their $\u1^{\otimes m}$ \reps are reducible.
Reducible and irreducible degenerate representations have
the same quantum numbers of fractional charge, spin and statistics.
The existing experiments at hierarchical filling fractions were
sensible to these data only; therefore their successful interpretation
within the chiral-boson theory is consistent with our theory.
More refined experiments are needed to test the difference.


\bigskip
\noindent{\bf The $\winf$ Minimal Models}

Degenerate \reps are common in conformal field theory:
if the central charge and the weight of a given representation satisfy
certain algebraic relations, some of its states decouple, and should be
projected out to obtain an irreducible representation.
A general fact is that the fusion of
degenerate \reps only gives degenerate \reps of the same type; thus
it is possible to find sets of degenerate \reps which are closed
under the fusion rules; these build the minimal models \cite{bpz}.
There are specific minimal models for any symmetry algebra:
the well-known ones are the $c<1$ Virasoro minimal models;
larger symmetry algebras, like $\winf$, have $c>1$ minimal models \cite{bpz}.
The minimal models have less states than the generic theories with
the same symmetry, due to the projection; for the same reason, they
have a richer dynamics.

The $\winf$ minimal models are {\it not} realised by the multi-component chiral
boson theories with $\widehat{U(1)}^{\otimes m}$ symmetry, because the latter
do not incorporate the projection for making irreducible the $\winf$
representations of degenerate type.
They are instead realised by the
$\u1\otimes {\cal W}_m(p=\infty)$ conformal theories \cite{kac2}, where the
${\cal W}_m(p)$ are the Zamolodchikov-Fateev-Lykyanov models with
$c=(m-1)\left[1-m(m+1)/p(p+1)\right]$ \cite{fz}.
We have found the minimal set of representations which are closed
under the fusion rules of these models, which is again a lattice
(\ref{latt}) satisfying special conditions, which makes it similar to
the weight lattice of the $SU(m)\otimes U(1)$ Lie algebra \cite{ctz5}.
We have also identified the physical charge of the excitations and the
Hall conductivity with arguments analogous to the one described before
in the chiral boson theory.
We have obtained the spectrum
\barr
\nu& =& {m\over mp \pm 1}\ , \qquad p >0 \ {\rm even}\ ,\qquad c = m\ ,
\nonumber\\
Q & = & {1\over pm \pm 1}\ \sum_{i=1}^m n_i \ ,\qquad
n_1 \ge n_2 \ge \cdots \ge n_m\ , \quad n_i \in {\bf Z}\ ,\nonumber\\
{ \theta\over\pi} &=& \pm\left( \sum_{i=1}^m n^2_i -
{p\over mp \pm 1}\left( \sum_{i=1}^m n_i \right)^2 \right) \ .
\label{wspec}\earr
These spectra agree with the experimental data and match the results
of the lowest-order Jain hierarchy (\ref{jspec})\footnote{
Note, however, the reduced multiplicities of eq.((\ref{wspec}).}
discussed in section 3.

This result has far-reaching consequences, both theoretical and experimental.
The physical mechanism which stabilizes the observed quantum Hall fluids
has both short and long distance manifestations.
At the microscopic level, it can be described by the Jain composite-electron
picture and by the size of the gaps;
in the scaling limit, by the minimality of the $\winf$ edge theory.
Actually, we find it rather natural that the theories with a minimal set
of excitations are also dynamically more stable.
This long-distance stability principle leads to a logically self-contained
edge theory of the fractional Hall effect:
a thorough derivation of experimental
results is obtained from the principle of $\winf$ symmetry, which is
the basic property of the Laughlin incompressible fluid.
This independent hierarchical construction is the main result of our
approach.

Furthermore, the detailed predictions of the $\winf$ minimal theories
are different from those of the chiral boson theories.
The main differences are as follows:

i) There is a {\it single} Abelian current, instead of $m$ independent
ones, and therefore a single elementary (fractionally) charged excitation;
there are neutral excitations, but they cannot
be associated to $(m-1)$ independent edges.

ii) The dynamics of these neutral excitations is new: they have
associated an $SU(m)$ (not $\widehat{SU(m)}_1$) ``isospin'' quantum number,
which is given by the highest weight \cite{ctz5},
\beq
{\bf \Lambda}=\sum_{a=1}^{m-1}\ {\bf \Lambda}^{(a)}\ \left(
n_a - n_{a+1} \right) \ ,
\label{sumwei}\eeq
where ${\bf \Lambda}^{(a)}$ are the fundamental weights of $SU(m)$ \cite{wyb}
and $\{n_i\}$ are the integer labels of eq.(\ref{wspec}).
More precisely, they are associated to
${\cal W}_m(p=\infty)$ \reps, whose fusion rules are isomorphic to
the branching rules of the $SU(m)$ Lie algebra.
Therefore, the neutral excitations are quark-like and their
statistics is non-Abelian.
For example, the edge excitation corresponding to the electron is composed of
them, and carries both the additive electric charge and the $SU(m)$ isospin.

iii) The degeneracy of particle-hole excitations at fixed angular momentum
is modified by the projection of the minimal models. This counting of states
is provided by the characters of degenerate $\winf$ representations,
which are known \cite{kac2}.
If the neutral $SU(m)$ excitations have a bulk gap, the particle-hole
degeneracy of the ground state (the Wen topological order on the disk
\cite{topord}) is different from the corresponding one of $\u1^{\otimes m}$
excitations.
This can be tested in numerical diagonalizations of few electron systems.

\vbox to .5 in {\vfill}

\noindent{\bf Non-Abelian fusion rules and non-Abelian statistics}

The physical electron is identified
as the minimal set of $\winf$ \reps with unit charge and
integer statistics relative to all excitations in eq.(\ref{wspec}) \cite{ctz5}.
These conditions are fulfilled by a composite edge excitation
$n_i=(1+p,p,\dots,p)$,
which is made of $(mp)$ elementary charged {\it anyons} and the
{\it quark} elementary neutral excitation, i.e. the fundamental
$SU(m)$ isospin \rep, due to $(n_i-n_{i+1})=\delta_{i,1}$ in (\ref{sumwei}),
for filling fraction $\nu=m/(mp \pm 1)$.

A conduction experiment which could show the composite nature of
the electron has been proposed \cite{kane}.
It is a modification of the ``time-domain'' experiment \cite{tdom},
in which a very fast electric pulse was injected at the boundary
of a disk sample and a chiral wave was detected at another
boundary point. The proposed experiment will also detect the neutral
excitation in the electron, which propagates at a different speed.

The compositeness of the electron also plays a role in
the resonant tunnelling experiment \cite{milli}, in which
two edges of the sample are pinched at one point, such that the
corresponding edge excitations, having opposite chiralities, can interact.
At $\nu=1/3$, the point interaction of two elementary anyons
is relevant and determines the {\it scaling law} $T^{2/3}$ for
the conductance \cite{tunn}.
This scaling of the tunnelling resonance peaks is verified experimentally.
On the other hand, off-resonance and at low temperature, the conductance
is given by the tunnelling of the whole electron, with a different
scaling law in temperature \cite{wen}.

These experiments involve processes with one or two electrons:
their quark compositeness can be seen in four-electron processes,
like scattering.
Indeed, the expansion of the four-point function of the
electrons in intermediate channels is determined by the
fusion ($SU(2)$ isospin for $m=2$) of an electron pair. This is, schematically,
\beq
\langle\Omega| \Psi^\dagger(1)\Psi^\dagger(2)\Psi(3)\Psi(4) |\Omega\rangle
=\sum_{s=0,1} \langle\Omega| \Psi^\dagger(1)\Psi^\dagger(2) |\{s\}\rangle
\langle \{s\}| \Psi(3)\Psi(4)|\Omega\rangle\ ,
\label{nonab}\eeq
where the two channels follow from the addition of two one-half isospin values.
More than one intermediate channel are also created in the
adiabatic transport of two electrons around each other, in presence of
two other excitations, because the amplitude for this process is again a
four-point function. For generic excitations, the monodromy phases form a
matrix, which gives a non-Abelian
representation of the braid group \cite{wilc}.
This is precisely the notion of non-Abelian statistics\footnote{
For a general discussion of non-Abelian statistics in the quantum Hall
effect, see ref.\cite{moore}.}.
These monodromy properties also determine the degeneracy of
the ground state on a torus geometry, the so-called topological order
\cite{wen}. This depends on the type of the \reps carried
by the edge excitations \cite{bpz}, and has not yet been computed for the
$\u1\otimes{\cal W}_m$ ones.

\bigskip

\noindent{\bf The degeneracy of excitations above the ground state}

In order to discuss this point, we must rewrite the spectrum (\ref{wspec}).
Let us consider, for example, the $m=2$ chiral theories, which are relevant
for $\nu=2/5, \dots$.
The corresponding $\winf$ minimal model
is constructed from degenerate representations of the type
$\widehat{U(1)}\otimes {\cal W}_2$,
where the ${\cal W}_2$ algebra is the $c=1$ Virasoro algebra.
As explained in \cite{ctz5}, these degenerate Virasoro representations
carry an $SU(2)$ isospin quantum number, as required by the fusion
rules \cite{bpz}.
Consider any excitation $(n_1,n_2)$ associated to a $\u1\otimes{\rm Vir}$
\rep, labelled by the $\u1$ charge $Q\propto (n_1+n_2)$ and the
$SU(2)$ isospin $s=|n_1-n_2|/2$.
Divide the square lattice $(n_1,n_2)$
into charged excitations and their neutral daughter excitations by introducing
the change of integer variables $(n_1,n_2) \to (l, n)$:
\beq
{\rm I}:\left\{
\begin{array}{l l}
2 l & =n_1+n_2 \\ 2n & =n_1-n_2 > 0 \\
& \ \ (n_1+n_2 \ {\rm even}), \end{array} \right.
\qquad\quad
{\rm II}:\left\{
\begin{array}{l l}
2 l +1& =n_1+n_2 \\ 2n +1& =n_1-n_2 > 0 \\
& \ \ (n_1+n_2 \ {\rm odd}). \end{array} \right.
\label{intcha}\eeq
The spectrum (\ref{wspec}) can be rewritten, for $\nu=2/(2p+1)$,
\beq
{\rm I}:\left\{
\begin{array}{l l}
Q & = {2l \over 2p+1}\ , \\
{1\over 2}{\theta\over \pi} & = {1\over 2p+1} l^2 +n^2  \end{array} \right.
\qquad\qquad
{\rm II}:\left\{
\begin{array}{l l}
Q & = {2 \over 2p+1}\left( l +{1\over 2} \right) \\
{1\over 2}{\theta\over \pi} & =
{1\over 2p+1} \left( l+{1\over 2} \right)^2 +{(2n+1)^2 \over 4}
\end{array} \right. \ .
\label{splitsp}\eeq
The $\u1$ charged excitations have  the same spectrum
$Q=\nu k,\ \theta/\pi=\nu k^2 $, of the simpler Laughlin fluids $(m=1)$.
Moreover, the infinite tower of neutral daughters ($n>0$) are characterized
by the conformal dimensions $h=(2n)^2/4\ $ (resp. $h=(2n+1)^2/4$).

The number of excitations above the ground state depends
on whether the neutral excitations have a bulk gap or not.
This affects also the thermodynamic quantities like the specific heat.

As said before, the charged edge excitations correspond to Laughlin
quasi-particles vortices in the bulk of the incompressible fluid,
which spill their density excess or defect to the boundary.
They have an (non-universal) gap proportional to the electrostatic energy of
the vortex core, which is not accounted for by the edge theory \cite{laugh}.
On the other hand, the bulk excitations corresponding to {\it neutral} edge
excitations are not well understood yet.
If they have a gap, they could exhibit the internal structure of the
quasi-particle vortex, or be bound states of a quasi-particle and a
quasi-hole; these would be localized two-dimensional excitations.
Neutral and charged gapful excitations can be thought of as analogs of the
{\it breathers} and {\it solitons} of one-dimensional integrable models,
respectively, \cite{inte}.
On the other hand, gapless neutral excitations would be pure effects
of the structured edge.

In the gapful case, the excitations above the ground state are the
particle-hole excitations (\ref{neut}), which are described by the $\winf$
\rep $(n=l=0)$ in (\ref{splitsp}).
The degeneracy of these states is encoded in the character of the \rep
\cite{bpz}.
In the gapless case, the states contained in the neutral
daughter Virasoro representations ($n>0\ ,\ l=0$) also contribute to these
degeneracies,
because they have integer spin (Virasoro dimension) and are indistinguishable.
Actually, the infinite tower of Virasoro \reps
($n>0\ ,\ l$ fixed) of each charged parent state $(l,n=0)$
can be summed (with their multiplicity one) into a single $\u1$ \rep
\cite{ctz5}.
In this case, the $m=2$ $\winf$ square-lattice spectrum (\ref{splitsp})
reduces to a one-dimensional array of $\u1^{\otimes 2}$ \reps,
with spectrum
\beq
{\rm I}: \ {1\over 2}{\theta\over\pi} ={1\over 2p+1} l^2\ ,\qquad\qquad
{\rm II}: \ {1\over 2}{\theta\over\pi} =
{1\over 2p+1} \left(l+{1\over 2} \right)^2 + {1\over 4}\ ,
\label{triv}\eeq
where the second $\u1$ eigenvalue is not observable.

We can repeat this resummation for the corresponding chiral boson theory
of the Jain fluid \cite{ctz5}.
The spectrum of charge and fractional statistics is
again given by (\ref{splitsp}), with multiplicities given by
$n\in {\rm Z}$: each $(l,n)$ value corresponds to a $\u1\otimes\u1$
\rep now. If they are gapless, the neutral daughter
$\u1\otimes\u1$ representations $((l,n),\ n\neq 0\in {\bf Z})$ of
each charged \rep $(l,0)$ can be similarly summed up into
one \rep of the larger algebra $\u1\otimes\widehat{SU(2)}_1$,
the non-Abelian current algebra of level one \cite{bpz}\cite{kmat}.
The spectrum of $\u1\otimes\widehat{SU(2)}_1$ representations is again
given by (\ref{triv}).

We can now compare the predictions of the $\winf$ minimal models and
the chiral boson theories for the degeneracy of the excitations above the
ground state.
This degeneracy can be measured in numerical simulations of a few electron
system in the disk geometry, by charting the eigenstates
of the Hamiltonian below the bulk gap \cite{wen}\cite{topord}.
Consider, for example, the $\nu=2/5$
$\ (m=p=2)\ $ ground state ($(l=0,n=0)$ in ((\ref{splitsp})).
In the following table, we report the degeneracies encoded in the
characters of the relevant $\u1\otimes {\rm Vir}$,
$\ \u1^{\otimes 2}$ and $\u1\otimes\widehat{SU(2)}_1$ \reps \cite{ctz5}:
\beq
\begin{array}{c | r r r r r r}
 \Delta J                   & 0 & 1 & 2 & 3  & 4  & 5  \\ \hline
 \u1\otimes {\rm Vir}       & 1 & 1 & 3 & 5  & 10 & 16  \\
 \u1\otimes\u1              & 1 & 2 & 5 & 10 & 20 & 36  \\
 \u1\otimes\widehat{SU(2)}_1& 1 & 4 & 9 & 20 & 42 & 80
\end{array}
\label{degtab}\eeq

As said before, if neutral daughter excitations are gapful, they should
not be counted,  and the degeneracy is only given by the particle-hole
excitations encoded in the ground state character of each theory.
On the other hand, if they are gapless, the total
degeneracy is given by the characters of the resummed \reps \cite{ctz5}.
We conclude that:

i) The observation of $\u1\otimes {\rm Vir}$ degeneracies confirms
the $\winf$ minimal theory with gapful neutral excitations;

ii) The $\u1\otimes\u1$ degeneracies are found both in the $\winf$ minimal
theory with gapless neutral excitations and in the chiral boson theory with
gapful ones;

iii) The $\u1\otimes\widehat{SU(2)}_1$ degeneracies support the chiral boson
theory with gapless neutral excitations.

Numerical results known to us \cite{wen} are not accurate
enough to see the differences in table ((\ref{degtab}).
Note the characteristic reduction of states of $\winf$ minimal models.

These remarks on the gap for neutral excitations do not affect
the previous discussion of the conduction experiments,
where excitations move along one edge or are transferred between two edges
at the same Fermi energy, such that bulk excitations are never produced.
Although the resummation of the neutral daughter ${\cal W}_m$ excitations gives
Abelian excitations, these are not $\winf$ irreducible, and thus unlikely to be
produced experimentally.
We think that only irreducible $\winf$ excitations, i.e.,
the elementary ones, can be naturally produced in a real system by an
external probe, for example by injecting an electron at the edge.

\bigskip

\noindent{\bf Remarks on the $SU(m)$ and $\widehat{SU(m)}_1$ symmetries}

We would like to explain the type of non-Abelian symmetry of
the $\winf$ minimal models
and clarify the differences with the chiral boson theories of the Jain
hierarchy, which have been also assigned the $SU(m)$ and $\widehat{SU(m)}_1$
symmetries \cite{read}\cite{kmat}\cite{froh}\cite{kane}.

Due to the $\u1\otimes{\cal W}_m$ construction of the $\winf$ models,
their excitations carry a quantum number which adds up as a $SU(m)$ isospin.
This {\it does not} imply that these models have the full $SU(m)$ symmetry,
in the usual sense of, say, the quark model of strong interactions,
because the states in each ${\cal W}_m$ representation do not form $SU(m)$
multiplets. As shown by the $m=2$ case, the quantum number $s=n/2$ of
Virasoro \reps is like the total isospin $S^2=s(s+1)$, but the $S_z$ component
is missing.
In some sense, the effects of the ${\cal W}_m$ non-Abelian fusion rules can
be thought of as a {\it hidden} $SU(m)$ symmetry.

On the other hand, it has been claimed that the chiral boson theories of the
Jain hierarchy  have a $SU(m)$ symmetry. The correct statement is, however,
that they possess $\u1\otimes\widehat{SU(m)}_1$ symmetry.
This means that their $\u1^{\otimes m}$ representations
can be rearranged into \reps of the $\u1\otimes\widehat{SU(m)}_1$
current algebra.
In the $\widehat{SU(m)}_k$ current algebra, the weights cannot be arbitrary,
but are cut-off by the {\it level} $k$  (e.g. for $m=2$, the spin $s$
can be $0\le s\le k/2$) \cite{bpz}. The level-one non-Abelian
current algebra has very elementary representations and their
fusion rules are made Abelian by this cut-off.

Therefore, the $\widehat{SU(m)}_1$ symmetry has no non-Abelian
physical effect, it is only a convenient reorganization of the
Abelian current algebra.
The non-Abelian character of the excitations is a characteristic
feature of the $\winf$ minimal models.

\section{Further developments}

In this paper, we have reviewed the $\winf$ theory of the edge excitations
in the Quantum Hall Effect.
In particular, we considered the {\it simplest} $\winf$ minimal models, which
are made of {\it one-congruence-class} degenerate representations only
\cite{kac2}. It would be interesting to generalize this construction,
in view of describing the experimentally observed filling fractions
$4/5,\ 5/7\ $, $4/11,\ 7/11,\ 8/11$, $\ 4/13,\ 8/13$, $9/13,\ 10/17$,
not explained here. Actually,
the $\winf$ minimal models can be generalized by considering two (or more)
congruence classes \cite{ctz5}.
There are analogies between this mathematical construction
and the Jain hierarchical construction of wave functions, which read
\beq
\Psi_\nu= D^{q/2}\ L^l\ D^{p/2}\ L^m\ {\bf 1}, \qquad\qquad p,q\ {\rm even},
\label{secord}\eeq
to second order of iteration \cite{jain}.
The number of fluids in any $\winf$ congruence
class corresponds to the number of Landau levels in (\ref{secord});
in both constructions, there are two independent elementary anyons,
each one accompanied by neutral excitations.
However, we have not yet proven a complete equivalence of the two
second-order hierarchies.

Another interesting development is the extension of the $\winf$
symmetry by additional degrees of freedom to the quantum incompressible
fluid, which might describe spinful electrons or multi-layer Hall devices.

\bigskip

\noindent{\bf Acknowledgements}

We would like to thank the organizers of the conference on
{\it Statistical Mechanics and Quantum Field Theory} and
the summer school on {\it Particles, Fields and Strings} for
succeeding in creating a stimulating atmosphere for discussions and
a positive interchange of ideas.
Throughout our project, we have greatly benefited from discussions
with V. Kac. We would also like to acknowledge the continuing
support of L. Alvarez-Gaum\'e and S. Fubini.
This work was supported in part by the CERN Theory Division,
the MIT Center for Theoretical Physics and the European Community program
``Human Capital and Mobility''.

\def\NP{{\it Nucl. Phys.\ }}
\def\PRL{{\it Phys. Rev. Lett.\ }}
\def\PL{{\it Phys. Lett.\ }}
\def\PR{{\it Phys. Rev.\ }}
\def\IJMP{{\it Int. J. Mod. Phys.\ }}
\def\MPL{{\it Mod. Phys. Lett.\ }}


\begin{thebibliography}{99}
\bibitem{prange} For a review see: R. A. Prange, S. M. Girvin, {\it The Quantum
		Hall Effect}, Springer Verlag, New York (1990).
\bibitem{laugh} R. B. Laughlin, \PRL {\bf 50} (1983) 1395;
		for a review see: R. B. Laughlin, {\it Elementary Theory: the
		Incompressible Quantum Fluid}, in \cite{prange}.
\bibitem{bpz}   A. A. Belavin, A. M. Polyakov and A. B. Zamolodchikov,
		\NP {\bf B 241} (1984) 333; for a review see:
		P. Ginsparg, {\it Applied Conformal Field Theory},
		in {\it Fields, Strings and Critical Phenomena},
		Les Houches School 1988, E. Brezin and J. Zinn-Justin eds.,
		North-Holland, Amsterdam (1990).
\bibitem{polch} See, for example: J. Polchinski, {\it ``Effective field
                theory and the Fermi surface''}, Lectures presented at
                TASI-92, Boulder (U.S.A.), June 1992, preprint
                NSF-ITP-92-132, UTTG-20-92, hep-th/9210046.
\bibitem{ctz1}  A. Cappelli, C. A. Trugenberger and G. R. Zemba,
		\NP {\bf 396 B} (1993) 465.
\bibitem{cdtz1} A. Cappelli, G. V. Dunne, C. A. Trugenberger and G. R.
		Zemba, \NP {\bf 398 B} (1993) 531.
\bibitem{ctz2}  A. Cappelli, C. A. Trugenberger and G. R. Zemba,
		\PL {\bf 306 B} (1993) 100.
\bibitem{cdtz2} A.Cappelli, G.V.Dunne, C.A.Trugenberger and G.R.Zemba,
		in the proceedings of {\it Common Trends in Condensed
		Matter and High Energy Physics}, Chia (Sardinia) 1992,
		L.Alvarez-Gaum\'e et al. eds., \NP {\bf B (Proc. Suppl.) 33C}
		(1993) 21.
\bibitem{ctz3}  A. Cappelli, C. A. Trugenberger and G. R. Zemba,
		\PRL {\bf 72} (1994) 1902.
\bibitem{ctz4}  A. Cappelli, C. A. Trugenberger and G. R. Zemba,
		{\it $\winf$ Dynamics of Edge Excitations in the
		Quantum Hall Effect}, preprint MPI-PhT/94-35, DFTT-16/94,
		cond-mat/9407095.
\bibitem{ctz5}  A. Cappelli, C. A. Trugenberger and G. R. Zemba,
		{\it Stable hierarchical Quantum Hall Fluids as
                $\winf$ minimal models}, preprint UGVA-DPT 1995/01-879
                DFTT-09/95, hep-th/9502021.
\bibitem{sakita}S. Iso, D. Karabali and B. Sakita,
		\NP {\bf B 388} (1992) 700, \PL {\bf B 296} (1992) 143.
\bibitem{flohr} M. Flohr and R. Varnhagen, {\it J. Phys.} {\bf A 27} (1994)
                3999; D. Karabali, \NP {\bf B 419} (1994) 437, {\bf B 428}
                (1994) 531.
\bibitem{shen}  I. Bakas, \PL {\bf B 228} (1989) 57;
		C. N. Pope, X. Shen and L. J. Romans, \NP {\bf B 339} (1990);
		for a review see: X. Shen, \IJMP  {\bf 7 A} (1992) 6953.
\bibitem{wen}   For a review, see: X. G. Wen, \IJMP {\bf 6 B} (1992) 1711.
\bibitem{stone} M. Stone, {\it Ann. Phys.} (NY) {\bf 207} (1991) 38,
		\PR{\bf 42 B} (1990) 8399, \IJMP  {\bf 5 B} (1991) 509.
\bibitem{wilc}  F. Wilczek, ed, {\it Fractional Statistics and Anyon
		Superconductivity} (World Scientific, Singapore, 1990).
\bibitem{gmp}   S. M. Girvin, A. H. MacDonald and P. M. Platzman,
                {\it Phys. Rev.} {\bf B 33 } (1986) 2206; S. M. Girvin,
                {\it "Collective Excitations"}, in \cite{prange}.
\bibitem{kac1}  V. Kac and A. Radul, {\it Comm. Math. Phys.} {\bf 157}
		(1993) 429; see also H. Awata, M. Fukuma, Y. Matsuo and
		S. Odake, {\it Representation theory of the $\winf$ Algebra},
		preprint hep-th/9408158.
\bibitem{kac2}  E. Frenkel, V. Kac, A. Radul and W. Wang, {\it
		${\cal W}_{1+\infty}\ $ and $\ {\cal W}(gl_N)\ $ with
		central charge $N$}, preprint hep-th/9405121.
\bibitem{radul} I. Vaysburd and A. Radul, \PL {\bf 274 B} (1992) 317.
\bibitem{hald}  F. D. M. Haldane, \PRL {\bf 51} (1983) 605;
		 B. I. Halpern, \PRL {\bf 52} (1984) 1583.
\bibitem{jain}  For a review see: J. K. Jain, {\it Adv. in Phys.}
		{\bf 41} (1992) 105.
\bibitem{hlr}   B. I. Halperin, P. A. Lee and N. Read, \PR {\bf B 47} (1993)
		7312.
\bibitem{nu1/2} R. R. Du, H. Stormer, D. C. Tsui, L. N. Pfeiffer and K. W.
West,
		\PRL {\bf 70} (1993) 2944; W. Kang, H. L. Stormer,
		L. N. Pfeiffer, K. W. Baldwin and K. W. West, \PRL {\bf 71}
		(1993) 3850.
\bibitem{halp}  B. I. Halperin, \PR {\bf B 25} (1982) 2185.
\bibitem{jtrans} X.- G. Wen, \MPL {\bf B 5} (1991) 39.
\bibitem{kmat}  J. Fr\"ohlich and A. Zee, \NP {\bf 364 B}
		(1991) 517; X.-G. Wen and A. Zee, \PR {\bf 46 B} (1993) 2290.
\bibitem{flo}   R. Floreanini and R. Jackiw, \PRL {\bf 59} (1987) 1873.
\bibitem{tilt}  L. W. Engel, S. W. Hwuang, T. Sajoto, D. C. Tsui and
		M. Shayegan, \PR {\bf B 45} (1992) 3418.
\bibitem{froh}  J. Fr\"ohlich, U. M. Studer and E. Thiran, {\it An ADE-O
		Classification of Minimal Incompressible Quantum Hall Fluids},
		preprint cond-mat/9406009; J. Fr\"olich et al., {\it The
		Fractional Quantum Hall Effect, Chern-Simons Theory, and
		Integral Lattices}, preprint ETH-TH/94-18,
		Cern SCAN-9409027.
\bibitem{read}  N. Read, \PRL {\bf 65} (1990) 1502.
\bibitem{tdom}  R. C. Ashoori, H. L. Stormer, L. N. Pfeiffer, K. W. Baldwin
		and K. West, \PR {\bf B 45} (1992) 3894;
\bibitem{milli} F. P. Milliken, C. P. Umbach and R. A. Webb, {\it Evidence
		for a Luttinger liquid in the Fractional Quantum Hall Effect},
		IBM preprint 1994.
\bibitem{tunn}  K. Moon, H. Yi, C. L. Kane, S. M. Girvin and M. P. A. Fisher,
		\PRL {\bf 71} (1993) 4381; P. Fendley, A. W. W. Ludwig and
		H. Saleur, {\it Exact Conductance through Point Contacts in
		the $\nu=1/3$ Fractional Quantum Hall Effect}, preprint
		cond-mat/9408068.
\bibitem{kane}  C. L. Kane and M. P. A. Fisher, {\it Impurity Scattering and
		Transport of Fractional Quantum Hall Edge States}, preprint
		cond-mat/9409028.
\bibitem{fz}    V. A. Fateev and A. B. Zamolodchikov, \NP {\bf B 280} (1987)
		644; V. A. Fateev and S. L. Lykyanov, \IJMP {\bf A 3} (1988)
		507.
\bibitem{wyb}   See for example: B. G. Wybourne, {\it Classical Groups for
		Physicists}, (Wiley, New York, 1974).
\bibitem{topord}X.- G. Wen, \PRL {\bf 70} (1993) 355.
\bibitem{moore} G. Moore and N. Read, \NP {\bf B 360} (1991) 362.
\bibitem{inte}  See for example: R. Rajaraman, {\it Solitons and Instantons},
		(North-Holland, Amsterdam, 1982).
\end{thebibliography}
\end{document}